\documentclass[aps,prl,manuscript,floats,showpacs,floatfix]{revtex4}

\usepackage{epsfig}
\usepackage{color}
\usepackage{bm}
\usepackage{latexsym}

\begin{document}

\newcommand{\hide}[1]{}
\newcommand{\earth}{\oplus}
\newcommand{\drsme}{\delta r_{SME}(t)}

\title{ {\bfseries Testing for Lorentz Violation:  Constraints on Standard-Model Extension Parameters via Lunar Laser Ranging} }

\author{James B. R. Battat, John F. Chandler, Christopher W. Stubbs}
\affiliation{Harvard-Smithsonian Center for Astrophysics, Cambridge, MA 02138}

\begin{abstract}
We present constraints on violations of Lorentz Invariance based on
Lunar Laser Ranging (LLR) data.  LLR measures the Earth-Moon
separation by timing the round-trip travel of light between the two
bodies, and is currently accurate to a few centimeters (parts in
$10^{11}$ of the total distance).  By analyzing archival LLR data
under the Standard-Model Extension (SME) framework, we derived six
observational constraints on dimensionless SME parameters that
describe potential Lorentz-violation.  We found no evidence for
Lorentz violation at the $10^{-6}$ to $10^{-11}$ level in these
parameters.
\end{abstract}

%
\maketitle


Lorentz symmetry, the idea that physical laws take the same form in
any inertial frame, irrespective of the orientation or velocity of the
frame, underpins the Standard Model and both Special and General
Relativity.  Attempts to quantize gravity have resulted in theories
that allow for violations of Lorentz symmetry \cite{amelinoCamelia2005}.  For
example, spontaneous Lorentz-symmetry breaking is possible in certain
string theories \cite{KOS_SAMUEL}.  At the Planck Energy, $\sqrt{\hbar
c^{5}/G}\sim10^{19}$ GeV, the Standard Model and General Relativity
(GR) have comparable influence \cite{SCHWARZ}.  Lorentz symmetry may
not hold in that regime.  Although no existing experiment can probe
these energies, a Lorentz symmetry violation may also manifest itself
at much lower energies where existing experiments do have sensitivity
\cite{CK1998}.  At present, no experiment has detected a violation of
Lorentz symmetry.

The Standard-Model Extension (SME) is a generalized effective field
theory that adds a complete set of Lorentz-violating terms to the
minimal standard model \cite{CK1998}.  The theory of the gravitational sector of the SME was developed in \cite{KOS2004} and \cite{BK2006}.  The SME provides a framework to
analyze and compare the results of Lorentz symmetry experiments just
as the Parametrized Post Newtonian (PPN) framework allows for the
direct comparison of tests of gravity \cite{WillNordtvedt1972}.
Recent calculations of the observable consequences of Lorentz symmetry
violations in the pure-gravity sector of the minimal SME \cite{BK2006}
showed that existing Lunar Laser Ranging (LLR) data would be sensitive
to a subset of the pure-gravity SME parameters.  LLR measures the time
of flight of photons between a ranging station on the Earth and
retro-reflectors on the lunar surface \cite{WTB2005}.  This experiment
has been ongoing since the Apollo astronauts landed on the lunar
surface in 1969.  Over the past 38 years the measurement precision has
improved by more than two orders of magnitude and currently LLR data
can be used to determine the orbit of the Moon around the Earth to a
few millimeters, or a few parts in $10^{12}$ of the total range
\cite{WTB2004,APOLLOINSTR}.  In this \emph{Letter}, we present
constraints on six linearly-independent combinations of pure-gravity
sector SME parameters using archival, centimeter-precision LLR data.

\begin{figure}
\includegraphics[width=\columnwidth,keepaspectratio,clip]{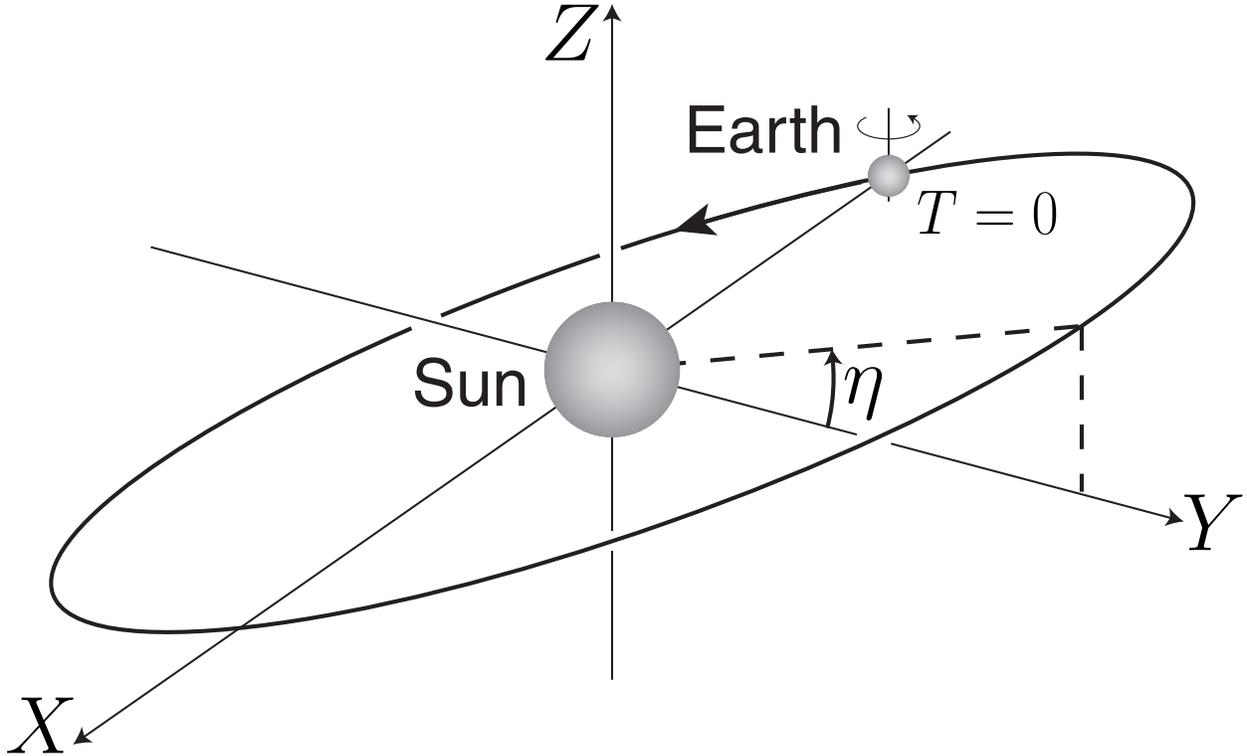}
\caption{\label{fig:celestialEquatorialReferenceFrame}The Sun-centered
celestial-equatorial plane.  The equator of the Earth defines the
XY-plane and the Earth's rotation axis is aligned with the Z-axis.
The orbital plane of the Earth (the ecliptic) is inclined to the
XY-plane by the obliquity angle, $\eta\approx23.5^{o}$.  The Earth is
shown at a descending node (a Vernal Equniox) which also defines the
origin of our time coordinate, $t$.  Reprinted figure with permission
from \cite{BK2006}.  Copyright 2006 by the American Physical Society.}
\end{figure}

A Lorentz violation would manifest itself as oscillatory perturbations
to the lunar orbit \cite{BK2006}.  In this section, we describe the
convention which we have adopted from \cite{BK2006} for our LLR data
analysis.  To explore these perturbations, it is convenient to work in
a Sun-centered celestial-equatorial frame, (see Figure
\ref{fig:celestialEquatorialReferenceFrame}).  In this system, the
Earth's equator defines the $XY$-plane and the Earth's spin angular
momentum vector points along the $+\hat{Z}$ direction.  The Earth
orbits the Sun in a plane (the ecliptic) which is inclined to the
$XY$-plane by an angle $\eta$ (approximately 23.5 degrees).  \hide{(in
celestial mechanics this angle is called the \emph{obliquity} and
often takes the symbol $\epsilon$).}  The Earth's orbit intersects the
$XY$-plane at two points, called the ascending and descending nodes.
In Figure \ref{fig:celestialEquatorialReferenceFrame}, the Earth is
pictured at the descending node, on the negative $X$-axis (note the
arrow that indicates the direction of the Earth's motion along its
orbit).  At this time, commonly called the Vernal Equinox, the Sun,
from the point of view of the Earth, ascends through the Earth's
equatorial plane.  The origin of $t$, our time coordinate (called $T$
in \cite{BK2006}), coincides with the presence of the Earth at the
descending node.

\begin{figure}
\includegraphics[scale=0.5,clip]{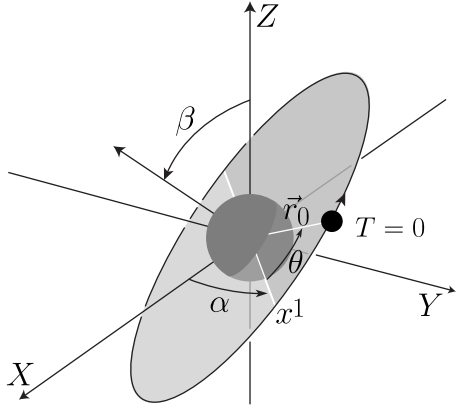}
\caption{\label{fig:lunarOrbitReferenceFrame}Lunar orbital parameters:
Here the Earth is shown translated to the center of the Sun-centered
coordinate system.  The lunar orbit is described by $r_{0}$, the mean
distance between the Earth and Moon, $e$ (not labeled) the
eccentricity of the lunar orbit, $\alpha$, the longitude of the
ascending node, $\beta$, the inclination of the lunar orbital plane
with respect to the Earth's equatorial plane and $\theta$, the angle,
along the Lunar orbit, subtended by the ascending node line and the
position of the Moon at $t=0$.  Reprinted figure with permission from
\cite{BK2006}.  Copyright 2006 by the American Physical Society.}
\end{figure}

Figure \ref{fig:lunarOrbitReferenceFrame} depicts the orbit of the
Moon about the Earth.  The longitude of the ascending node and the
inclination of the lunar orbit are labeled $\alpha$ and $\beta$,
respectively, and $r_{0}$ is the mean orbital radius of the Moon.  The
lunar orbital phase, $\theta$, corresponds to a unique reference frame
for an arbitrary choice of the $t=0$ Vernal Equinox.  The eccentricity
and longitude of perigee are not specified in this description.

In this coordinate system, Lorentz-violating perturbations to the
Earth-Moon separation can be expressed in the SME framework using the
following fourier series:

\begin{equation}
\drsme = 
   \sum_{n} 
     \left[ A_{n}\cos\left(\omega_{n} t + \phi_{n} \right) + 
            B_{n}\sin\left(\omega_{n} t + \phi_{n} \right)
     \right].
\label{eq:deltaR_SME}
\end{equation}

The dominant contributions to $\drsme$ occur at the following four
frequencies: $\omega$, $2\omega$, $2\omega-\omega_{0}$ and
$\Omega_{\earth}$.  Here (as in \cite{BK2006}) $\omega$ is the mean
lunar orbital (sidereal) frequency, $\omega_{0}$ is the anomalistic
lunar orbital frequency (perigee to perigee) and $\Omega_{\earth}$ is
the mean Earth orbital (sidereal) frequency.  The corresponding
amplitudes ($A_{n}$ and $B_{n}$) for these frequencies are listed in
Table \ref{table_smeRangePerturbations}.  The dominant perturbations
to the lunar orbit are controlled by a set of six linearly-independent
combinations of Lorentz-violating SME parameters,
$\bar{s}_{LLR}=\left\{ \left(\bar{s}^{11}-\bar{s}^{22}\right),
                       \bar{s}^{12},\bar{s}^{02},\bar{s}^{01},
                       \bar{s}_{\Omega_{\earth},c},
                       \bar{s}_{\Omega_{\earth},s}
               \right\}$ 
(see \cite{BK2006} for a description of these parameters).

At present the only published constraints on the gravitational sector
SME parameters come from measurements of the perihelion shifts of
Mercury and the Earth and from an argument based on the current
alignment of the solar spin axis and the angular momentum vector of
the planetary orbits \cite{BK2006}.  This paper provides the first LLR
constraints on SME parameters.

\begin{table}
\begin{tabular}{llr}
\hline 
& Amplitudes ($A_n$, $B_n$)& Phases ($\phi_n$) \\
\hline
$A_{2\omega}$ & $-\frac{1}{12}\left(\bar{s}^{11}-\bar{s}^{22}\right)r_{0}$ & $2\theta$\\
$B_{2\omega}$ & $-\frac{1}{6}\bar{s}^{12}r_{0}$                            & $2\theta$\\
$A_{2\omega-\omega_{0}}$ & $-\omega er_{0}\left(\bar{s}^{11}-\bar{s}^{22}\right)/16\left(\omega-\omega_{0}\right)$ & $2\theta$\\
$B_{2\omega-\omega_{0}}$ & $-\omega er_{0}\bar{s}^{12}/8\left(\omega-\omega_{0}\right)$ & $2\theta$\\
$A_{\omega}$ & $-\omega\left(\delta m\right)v_{0}r_{0}\bar{s}^{02}/M\left(\omega-\omega_{0}\right)$ & $\theta$\\
$B_{\omega}$ & $\omega\left(\delta m\right)v_{0}r_{0}\bar{s}^{01}/M\left(\omega-\omega_{0}\right)$  & $\theta$\\
$A_{\Omega_{\earth}}$ & $V_{\earth}r_{0}\left(b_{1}/b_{2}\right)\bar{s}_{\Omega_{\earth}c}$ & $0$\\
$B_{\Omega_{\earth}}$ & $V_{\earth}r_{0}\left(b_{1}/b_{2}\right)\bar{s}_{\Omega_{\earth}s}$ & $0$\\
\hline
\end{tabular}
\caption{\label{table_smeRangePerturbations} The leading-order
amplitudes from Equation \ref{eq:deltaR_SME} and their associated
phases (from \cite{BK2006}).  $M$ and $\delta m$ are the sum and
difference of the Earth and Moon masses, while $V_\earth$ and $v_0$
are the mean Earth and Moon orbital velocities, respectively,
normalized to the speed of light.  The unitless parameters $b_1$ and
$b_2$ are functions of $\eta$, $\alpha$, $\beta$, $r_0$,
$\Omega_\earth$ and the Earth's quadrupole moment.}
\end{table}

\begin{table}
\begin{tabular}{l}
\hline 
SME Parameter Partial Derivatives\\
\hline
$\frac{\partial\delta r}{\partial\left(\bar{s}^{11}-\bar{s}^{22}\right)} =
  -\frac{r_{0}}{12}\cos\left(2\omega t+2\theta\right) - 
   \frac{\omega er_{0}}{16\left(\omega-\omega_{0}\right)}\cos\left[\left(2\omega-\omega_{0}\right)t+2\theta\right]$\\
$\partial\delta r/\partial\bar{s}^{12} =
  -\frac{r_{0}}{6}\sin\left(2\omega t+2\theta\right) -
   \frac{\omega er_{0}}{8\left(\omega-\omega_{0}\right)}\sin\left[\left(2\omega-\omega_{0}\right)t+2\theta\right]$\\
$\partial\delta r/\partial\bar{s}^{02} = 
  -\frac{\omega\left(\delta m\right)v_{0}r_{0}}{M\left(\omega-\omega_{0}\right)}\cos\left(\omega t+\theta\right)$\\
$\partial\delta r/\partial\bar{s}^{01} = 
   \frac{\omega\left(\delta m\right)v_{0}r_{0}}{M\left(\omega-\omega_{0}\right)}\sin\left(\omega t+\theta\right)$\\
$\partial\delta r/\partial\bar{s}_{\Omega_{\earth}c} = 
   V_{\earth}r_{0}\left(\frac{b_{1}}{b_{2}}\right)\cos\left(\Omega_{\earth}t\right)$\\
$\partial\delta r/\partial\bar{s}_{\Omega_{\earth}s} = 
   V_{\earth}r_{0}\left(\frac{b_{1}}{b_{2}}\right)\sin\left(\Omega_{\earth}t\right)$\\
\hline
\end{tabular}
\caption{\label{table_smePartialDerivatives} The partial derivatives of the
SME perturbation to the lunar range with respect to each $\bar{s}_{LLR}$
parameter.}
\end{table}


The fundamental LLR observable is the travel time of light (usually a
pulsed laser beam) that is propagated from a transmit station on the
Earth to retroreflector arrays on the Lunar surface and is reflected
back to a receive station on the Earth.  Typically the receive station
is the same as the transmit station.  The return signal from the Moon
is weak -- typically 1 to 5 photons per minute for the data used here
-- and single-photon counters are requred to detect and time-tag lunar
reflection events.

LLR data have historically been presented as ``normal points'' which
combine a series of single-photon lunar reflection events to achieve a
higher signal-to-noise ratio measurement of the lunar range at some
characteristic time for that data series.  In this analysis, we used
14,401 normal points spanning September 1969 through December 2003,
taken from the public LLR data archive \cite{ILRS}.  The bulk of these normal points were generated by the McDonald Laser Ranging Station
in Texas, USA \cite{mlrs} and the Observatoire de la C\^{o}te d'Azur station in Grasse, France \cite{oca}.
A single normal point is typically generated from an observation spanning 5 to 20
minutes during which anywhere from a few to a hundred photons are
collected.  The lunar ranges reported in these normal points
incorporate station-specific hardware corrections.


In order to analyze LLR data, one relies on detailed models of the
solar system ephemeris and dynamics.  To our knowledge, there are only
a few such models in existence capable of this analysis
\cite{MUELLER1997,WTB2005,PEP_description}.  None of these codes
explicitly include the SME framework in their equations of motion for
solar system bodies.  Nevertheless it is still possible to constrain
the $\bar{s}_{LLR}$ parameters.

We used the Planetary Ephemeris Program (PEP) \cite{PEP_description}
to extract SME parameter constraints from LLR observations.  PEP uses
its ephemeris and dynamics model, along with an inital set of model
parameters, to compute a range prediction and the partial derivatives
of range with respect to each model parameter at the time of each
normal point.  It then performs a weighted, linear least-squares
analysis to calculate adjustments to the model parameters in order to
minimize the difference between the observations and the model.  Were
the lunar range model linear in all parameters, this method would be
exact.

If one is concerned about non-linearities, one can solve for model
parameters and then re-integrate the equations of motion, iterating
until the parameter estimates converge.  Over the past several decades
the traditional ({\em i.e.} non-SME) analyses have done just that,
resulting in agreement between model and data at the few centimeter
level.  Current model parameter values are therefore highly refined,
and the weighted least-squares analysis sits firmly in the linear
regime.  As a result, it is not necessary to iterate when estimating
new model parameters.  Because the lunar range model is linear in the
$\bar{s}_{LLR}$ parameters (see Equation \ref{eq:deltaR_SME} and Table
\ref{table_smeRangePerturbations}), the inclusion of these parameters
in the analysis preserves linearity, as confirmed by the small
adjustments to non-SME parameters seen in our solutions.  Performing
an iterative solution for SME parameters requires the inclusion of the
SME terms in the equations of motion.

We computed the partial derivatives of lunar range with respect to
each $\bar{s}_{LLR}$ parameter and provided this information to PEP
prior to solving for the best-fit parameter adjustments.  This
approach is equivalent to explicitly including the $\bar{s}_{LLR}$
parameters in the equations of motion and setting their \emph{a
priori} values to zero (in which case $\drsme=0$ so there is no SME
contribution to the lunar orbit).  We therefore treated any Lorentz
violation as a small perturbation to a known orbit.  The terms in the
covariance matrix quantify any ``crosstalk'' between SME parameters
and other fitted quantities.


The solar system is complex.  When modeling the expected light travel
time between an LLR station on the Earth and a reflector on the lunar
surface, one must account not only for the gravitational perturbations
from the eight planets and Pluto, but also those of planetary
satellites, asteroids, asphericities in the Sun, Earth and Moon, as
well as various relativistic and non-gravitational effects (for a more
complete description of relevant physical effects, see
\cite{WTB2005}).  As a result there are many hundreds of parameters
that have influence on the Earth-Moon range time.  Not surprisingly,
there are parameter degeneracies and LLR data alone cannot determine
all of these parameters.  Therefore LLR-only analyses suffer from
systematic uncertainties in model parameter estimates and the formal
error reported by the least-squares solution will underestimate the
true model parameter uncertainties.  These systematic uncertainties
can dominate the formal errors.  If auxiliary solar system data is
included in the analysis (e.g. planetary radar ranging) then the
number of model parameters grows and new parameter degeneracies arise.
Having chosen to perform our analysis using LLR data alone, we
accounted for the underestimation of parameter uncertainties by
inflating the formal parameter uncertainties from our least-squares
solution by a uniform factor which we label $F$.  This is numerically
equivalent to a uniform scaling of the uncertainty associated with
each normal point by $F$.

In order to determine the $F$ factor, we performed an analysis of the
LLR data in which we froze all $\bar{s}_{LLR}$ parameters at zero, but
allowed the PPN parameters, $\beta$ and $\gamma$ to vary.  We know
from other experiments \cite{BERTOTTI2003, VLBI2004, CHANDLER1996,
WTB2004} that $\beta$ and $\gamma$ are consistent with their GR value
of unity to within a part in $10^{3}$ or better.  We found that we
must scale our uncertainties on $\beta$ and $\gamma$ by $F=20$ to be
in accord with these earlier results.

\begin{figure}
\includegraphics[width=\columnwidth,keepaspectratio,clip]{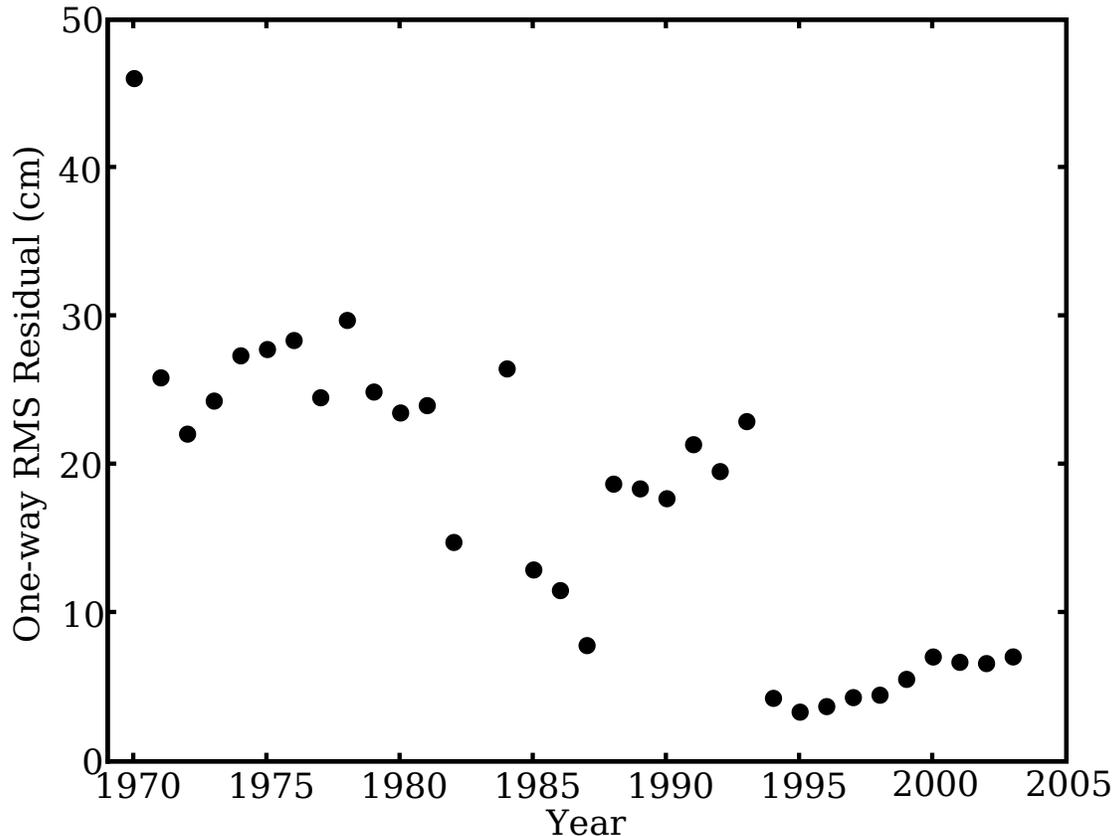}
\caption{\label{rmsResiduals} The annual RMS residual between the LLR
data and our best-fit model for the lunar range.  The residual RMS in
1969 is over 300 cm.  We omitted this point from the plot for clarity,
but the two data points from that year were included in the analysis.
Over this timespan, the potential Lorentz-violating signals would have
undergone at least 34 cycles.}
\end{figure}

We then estimated the values of the set of $\bar{s}_{LLR}$ parameters
while holding the PPN parameters at their GR values
($\beta=\gamma=1$).  The resulting parameter values and their
realistic errors (the formal errors scaled by the $F=20$ factor) are
reported in Table \ref{table_smeParamConstraints}.  All SME parameters
are within one and a half standard deviations of zero and are within
an order of magnitude of the sensitivity expected from 1 cm precision
LLR data as estimated in \cite{BK2006}.  There is no evidence for
Lorentz violation in the lunar orbit.  The model-data agreement,
binned annually, is shown in Figure \ref{rmsResiduals}.

\begin{table}
\begin{tabular}{lcr}
\hline 
Parameter & Predicted Sensitivity & This work \\
\hline
$\bar{s}^{11}-\bar{s}^{22}$  & $10^{-10}$ & $\left( 1.3 \pm 0.9 \right) \times 10^{-10}$\\
$\bar{s}^{12}$               & $10^{-11}$ & $\left( 6.9 \pm 4.5 \right) \times 10^{-11}$\\
$\bar{s}^{02}$               & $10^{-7}$  & $\left(-5.2 \pm 4.8 \right) \times 10^{-07}$\\
$\bar{s}^{01}$               & $10^{-7}$  & $\left(-0.8 \pm 1.1 \right) \times 10^{-06}$\\
$\bar{s}_{\Omega_{\earth}c}$ & $10^{-7}$  & $\left( 0.2 \pm 3.9 \right) \times 10^{-07}$\\
$\bar{s}_{\Omega_{\earth}s}$ & $10^{-7}$  & $\left(-1.3 \pm 4.1 \right) \times 10^{-07}$\\
\hline
\end{tabular}
\caption{\label{table_smeParamConstraints} The predicted sensitivity to each $\bar{s}_{LLR}$ parameter (from \cite{BK2006}) and the values derived in this work including the realistic (scaled)
uncertainties ($F\sigma$) with $F=20$.  In this analysis, the PPN
parameters were fixed at their GR values.  The SME parameters are all
within 1.5 standard deviations of zero.  We see no evidence for
Lorentz violation under the SME framework.}
\end{table}

To ensure that our analysis is sensitive to the signature of a Lorentz
violation in the lunar range, we generated a perturbed LLR dataset
with a hand-inserted Lorentz-violation signal corresponding to a value
of $\bar{s}^{11}-\bar{s}^{22}=9\times 10^{-10}$ (a factor of 10 larger
than the uncertainty for this parameter).  Our analysis recovers this
signal.  We found $\bar{s}^{11}-\bar{s}^{22}=\left(10 \pm 0.9 \right)
\times 10^{-10}$ with the other $\bar{s}_{LLR}$ parameters unchanged.

In conclusion, we have analyzed over 34 years of archival LLR data
using PEP to derive constraints (see Table
\ref{table_smeParamConstraints}) on six linearly-independent
Lorentz-violation parameters in the pure-gravity sector of the SME.
These are the first LLR-based SME constraints.  Using the same LLR
dataset, these constraints could be improved by performing a
simultaneous fit with auxiliary solar system data to help break
parameter degeneracies and thereby reduce the systematic error budget.
Based on the solutions for $\beta$ and $\gamma$ obtained in this work,
we estimate that the SME parameter constraints could be improved by a
factor of 5-10 with a joint fit.  In addition, we are involved in the
Apache Point Observatory Lunar Laser-ranging Operation project
(APOLLO), a next-generation LLR station that is currently collecting
millimeter-precision lunar range data \cite{APOLLOINSTR}.  We plan to
improve the SME parameter constraints by incorporating APOLLO data
into our analysis.

This project developed out of conversations with several colleagues
including E. Adelberger, Q. Bailey, J. Davis, A. Kosteleck{\'y},
J. Moran, T. Murphy, I. Shapiro and M. Zaldarriaga.  We would also
like to thank the LLR observers and the ILRS for the data collection
and normal point distribution.  J. B. R. B. acknowledges financial
support from the ASEE NDSEGF, the NSF GRFP and Harvard University.
Generous financial support was provided by the National Science
Foundation (Grant No. PHY-0602507).



\end{document}